\documentclass[10pt]{iopart}\usepackage{iopams}  %iop %for arxiv
\usepackage{graphicx}

\usepackage{mathrsfs} %{\mathscr{}}''''''''''
\usepackage{eucal} %jxzj
\usepackage{amssymb}%nova org
\usepackage{amsfonts}%nova org
\usepackage{amsbsy}%jxzj uncommond

\begin{document}

\title[]{Inference of the Universal Constancy of  Planck Constant based on First Principles}

\author{J.X. Zheng-Johansson}

\address{Institute of Fundamental Physics Research, 611 93 Nyk\"oping, Sweden}
 \address{(Jan. 15, 2013)}
%\date{April, 2007}
%\date{December 17, 2006}
%\address{July, 2007}
%\address{November 7, 2007}
%\address{
     %Dec 10,  2007; Nov 17, 2009; April 23-26,July 8th, Aug 05, Aug. 15, 2010; Aug 30, 2011, %Nov 07, Nov 23, 2011; Jan 30, Feb 10, July 2, 2012, Sept 4, 2012 Nov 1, 2012; Feb. 15, 2013 }

%\ead{jxzj@iofpr.org}

%\maketitle
%lesssim, gtrsim:  $\lesssim, \gtrsim$

\def\psi{\Psim}
\def\nt{\mathfrak{n}}
\def\subDN{{\mbox{\tiny{$\f N$}}}}
\def\Dcal{\mathcal{D}}
\def\citePlanck1900a{[1a]}
\def\citePlanck1900b{[1b]}
\def\Hcal{{\mathcal{H}}}
\def\f{D}
\def\b{b}
\def\M{M}
\def\P{P}
\def\Hscr{\mathscr{H}}

\def\Rbb{\mathbb{G}}
\def\Kbb{\mathbb{B}}
\def\Bbb{\mathbb{B}}

\def\R{r_{max}}
\def\Pscr{\mathscr{P}}
\def\Dscr{\mathscr{D}}

\def\Ds{\mathbb{D}}

\def\Xim{{\mit{\Xi}}}

\def\rw{\rightarrow}
\def\jm{{\j\mu}}
\def\kp{{\j'}}
\def\muk{{\n'}}
\def\p{{\mbox{\scriptsize{$+$}} \hspace{-0.03cm}}}
%$\p e$ $+e$
\def\pe{\p e}
\def\pq{\p q}
\def\minus{\mbox{-}}
\def\m{{\mbox{-}}}

\def\H{{a_{\Sigsub} \hspace{-0.05cm}}}
\def\arm{\mbox{\LARGE{\rm a}}}
 \def\Hn{I_{\n}}

\def\Ia{\mathcal{A}}
\def\hpbar{\abar}
\def\hpbars{\hpbar}
\def\hp{a}
               \def\abar{{a\hspace{-0.18cm}\mbox{{\small $^{_{_{-}}}$}}\hspace{-0.07cm}}}
                \def\abars{{\hspace{-0.03cm}a\hspace{-0.1cm}\mbox{\small{-}}\hspace{0.0cm}}}
\def\abar{{a\hspace{-0.18cm}\mbox{{\small $^{_{_{-}}}$}}\hspace{-0.07cm}}}
\def\abar{\overline{a}}
\def\Qcal{\mathbin{{Q}\mkern-8.5mu^{_{\mbox{\small{$\dash$}}}}\hspace{-0.04cm} }}

\def\Sa{{\mathfrak{S}}}
\def\nfrak{{\mathfrak{N}}}

\def\Ncal{\Ng}
\def\Ng{\Xim}

\def\Nb{\aleph}

\def\h{h}
\def\taubar{\mathbin{{\tau}\mkern-10.3mu_{^{{}^{{\mbox{\tiny{$-$}}}}\hspace{-0.10cm}}} }}
              %$\taubar$
           %\def\h{\eta}
\def\ho{\eta_0}
\def\hjn{\hp_{\jn}}

\def\Wst{{\mathbin{\Omega\mkern-4.1mu^{_{\mbox{\footnotesize{-}}}}}\hspace{-0.04cm}}}
\def\Wsts{{\mathbin{\Omega\mkern-4.2mu^{_{\mbox{\scriptsize{-}}}}}\hspace{-0.06cm}}}
\def\Wstsup{{\mathbin{\Omega\mkern-5.mu^{_{\mbox{\scriptsize{-}}}}}}}
          %\def\Wstt{{\mathbin{\Omega\mkern-3.5 mu^{_{\mbox{\scriptsize{-}}}}}}}  %--OK
          %\def\Wst{\mathbin{\Omega\mkern-6.mu^{_{\mbox{\scriptsize{-}}}}}}
         %$\Wst, \mit{\Omega}$ 

%\def\Nstat{{\mathcal{N}}}
%\def\Nstat{{\mathcal{N}}}

\def\Nst{{\Wst}}
\def\Nsts{{\Wsts}}
\def\Nstsup{{\Wstsup}}

\def\nsig{{\Sigsub\n}}
\def\nPi{{\n_{^{_\Pi}}}}

\def\pjn{p_{\j\n}}
\def\jn{{\j n}}
\def\xjn{{\j\n}}
\def\Ecal{{\mathcal{E}}}
         \def\Pcal{{\mathcal{P}}}
\def\Pcaln{\Pcal_{n}}
\def\Pcalbarn{ \overline{\Pcal_\n}}

\def\Pcalens{\Pcal_{\ens}}
\def\Pens{\Pcal_{\ens}}
\def\Pensm{\Pcal_{\ens,max}}
\def\Nstgm{\Nst_{i.g.m}}
\def\Nstam{\Nst_{i.a.m}}
\def\quadd{\ \ }
\def\la{\langle}
\def\ra{\rangle}
\def\Sigsub{{\mbox{{\tiny{$\Sigma$}}}}}
\def\Sig{\Sigma}
\def\bi{b^{i}}
\def\i{i}
\def\n{\nu}
\def\j{k}
\def\jj{{}}

\def\uscr{\mathscr{U}}
\def\uscrdotbar{\bar{\dot{\mathscr{U}}}}
\def\vac{{\rm{vac}}}
\def\Vcal{{\mathscr{V}}} %use with \usepackage{calrsfs} %jxzj?\usepackage{mathrsfs}
\def\Bsub{{\mbox{\small{$B$}}}}
\def\Nsub{{{\mbox{\tiny${N}$}}}}
\def\Hsub{{{\mbox{\tiny${H}$}}}}

\def\Hbar{\bar{H}}
\def\pbar{\bar{p}}

\def\exc{{\rm exc}}
\def\mini{0}
\def\bav{{\bar{b}}}
\def\v{{\rm v}}
\def\Hbar{\bar{H}}
\def\D{\Delta}
\def\bcal{b}
\def\bbar{\mathbin{{b}\mkern-9.5mu^{{\mbox{\tiny{$-$}}}}\hspace{-0.00cm} }}
\def\nstat{\nu}
\def\nst{\nu}
\def\engbar{\bar{\eng}}
\def\engobar{\bar{\eng}_0}
\def\psias{\psi}
\def\Phimas{\Phim}
\def\fas{f}
\def\rbb{\as}

\def\La{L}
\def\Ja{J}
\def\as{p}
\def\ioii{{\mbox{\normalsize${\frac{1}{2}}$}}}
\def\Rb{{\bf R}}
\def\rb{{\bf r}}
\def\ub{{\bf{u}}}
\def\hatu{\hat{u}_q}
\def\Nsub{{{\mbox{\tiny${N}$}}}}
\def\Pisub{{{\mbox{\tiny${\mit{\Pi}}$}}}}

\def\q{\bar{q}}
\def\xdot{\dot{x}}
\def\exc{{\rm ex}}
\def\ens{{ens}}
\def\Lcal{\mathcal{L}}
\def\Tcal{\mathcal{T}}
\def\Kcal{{\mathcal{K}}}
\def\Xcal{{\mathcal{X}}}

\def\Wvel{\Omegavel}
\def\Omegavel{\mathbin{{\mit\Omega}\mkern-13.mu^{_{\mbox{$-$}}}\hspace{-0.08cm}{}_d }}

\def\omegavel{{\w\mbox{\hspace{-0.38cm} \vspace{0.15cm}$-$\hspace{-0.02cm}}}}
\def\wvel{\omegavel_d}

\def\Ucal{\bar{\eng}_{0}}
\def\Omegavel{\mathbin{{\mit\Omega}\mkern-13.mu^{_{\mbox{$-$}}}\hspace{-0.08cm}{}_d }}
\def\Wvel{\Omegavel}

\def\q{\mathbin{q\mkern-11mu-}}
\def\PE{\mbox{\tiny{{\rm P.E.}}}}
\def\ME{\mbox{\tiny{{\rm M.E.}}}}
\def\QM{\mbox{\tiny{{\rm QM}}}}
\def\Psub{\mbox{\tiny{{\rm P}}}}
\def\Bsub{{\mbox{\tiny{{\rm B}}}}}

\def\ev{\epsilon}

\def\Ci{1}
\def\betamt{{\bf{b}}}
\def\kb{{\bar{k}}}
\def\kbf{{\bf{k}}}
\def\Kb{{\bf{K}}}
\def\cb{{\bf{c}}}

\def\pb{{\bar{p}}}
\def\pbf{{\bf{p}}}
\def\Acal{\mathscr{A}}
\def\Bcal{{I_{{\rm{ex}}}}}
\def\Ccal{{\cal{C}}}
\def\Vp{V}
\def\m{{{}_{\mbox{-}}}}
\def\Ccal{{\cal{C}}}
\def\p{{{}_{+\hspace{-0.1cm}}}}

\def\psipi{\psi_{\p}(1)}
\def\psipii{\psi_{\p}(2)}
\def\psimi{\psi_{\m}(1)}
\def\psimii{\psi_{\m}(2)}

\def\ai{\alpha(1)}
\def\aii{\alpha^{'}(2)}
\def\bi{\beta^{'}(1)}
\def\bii{\beta(2)}

\def\fa{f_r}
\def\fb{f_\ell}

\def\Ca{C_a}
\def\Cb{C_b}
\def\fbf{{\bf{f}}}
\def\Ocal{{\cal{O}}}
\def\psib{{\pmb{\psi}}}
\def\alphab{{\pmb{\alpha}}}
\def\sigmab{{\pmb{\sigma}}}
\def\sig{\sigma}
\def\Eb{{\bf E}}
\def\Bb{{\bf B}}
\def\ke{\kappa}
\def\nabb{{\pmb{\nabla}}}
\def\nablab{{\pmb{\nabla}}}
\def\vir{{\rm vir}}
\def\psitot{\psi}
\def\jb{{\bf{j}}}
\def\vel{v}
\def\velb{{\bf{v}}}

\def\Imtr{I}
\def\citeUnif{4?}
\def\App{}
\def\Qcal{{\mathcal{Q}}}
\def\Tcal{{\mathcal{T}}}
\def\Cross{Q}

\def\vphilim{f}
\def\ft{{\mathcal{B}}}
\def\vphibar{\mathbin{\varphi\mkern-12.5mu-}}
\def\vphi{\varphi}
\def\med{{\med}}
\def\Mcal{{\mathfrak{M}}}
\def\Sb{{\bf{S}}}
         \def\xia{{\mathcal{A}}}
\def\tha{\theta}

\def\nb{\bf{n}}
\def\zb{{\bf{z}}^0}
\def\phiv{\varphi}
\def\Lb{{\bf{L}}}
\def\velsub{_{\vel}}

\def\nablab{{\pmb{\nabla}}}
\def\velb{{\pmb{\vel}}}
\def\minus{\mbox{-}}

\def\Ab{{\bf{A}}_a}
\def\vel{\upsilon}
\def\Thm{\vartheta}
\def\Thetam{{\mit{\Theta}}}
\def\lb{{\bf l}}
\def\vb{{\bf{v}}}

\def\Rb{{\bf R}}
\def\pd{\partial}
\def\vphi{\varphi}

\def\psitot{\varphi}
\def\psiR{\widetilde{\psi}}
\def\psiL{\widetilde{\psi}^{{\rm vir}}}
\def\Phim{{\mit{\Phi}}}
\def\PhimR{\widetilde{ {\mit \Phi}}}
\def\PsimR{\widetilde{ {\mit \Psi}}}
\def\PsimL{{\widetilde{ {\mit \Psi}}}^{{\rm vir}}}
\def\a{\alpha}
\def\uav{\bar{u}}
\def\D{\Delta}
\def\th{\theta}
\def\r{{\mbox{\tiny${R}$}}}
\def\re{{\mbox{\tiny${R}$}}}
\def\Fmed{F_{{\rm a.med}}}
\def\med{{\rm med}}
\def\Lw{L_{\varphi}}
\def\Fb{{\bf{F}}}

\def\Efb{{\bf{E}}}
\def\Bfb{{\bf{B}}}
\def\Ac{ \varphi}
\def\Xsub{{\mbox{\tiny${X}$}}}
\def\Ysub{{\mbox{\tiny${Y}$}}}
\def\Zsub{{\mbox{\tiny${Z}$}}}
\def\Z{Z}
\def\Ksub{{\mbox{\tiny${K}$}}}
\def\W{{\mit \Omega}}
\def\Wd{\W_d{}}
\def\Nu{{\cal V}}
\def\Nud{\Nu_d{}}
\def\Eng{{\mathcal E}}
\def\Engvel{{\mathcal E}_\vel{}}
\def\Engvar{E}

\def\Engk{{\mathcal{K}}}
\def\Engvelk{{\mathcal{K}}_\vel{}}
\def\engvelk{\engks_\vel{}}
\def\engvel{\engs_\vel{}}
\def\eng{E}
\def\pvel{p_{_\vel}{}}
\def\Pvel{P_\vel{}}

\def\Kmscr{{\mathscr{K}}}
\def\Lscr{{\mathscr{L}}}
\def\engk{{K\hspace{-0.310cm}^{_{\mbox{-}}}\hspace{0.15cm}}}
\def\engqk{{\mathcal{K}}}
\def\engq{{\mathcal{E}}}
\def\Engq{{\mathcal{E}}}
\def\engq{\varepsilon}
\def\engqs{{\varepsilon}}
\def\engs{{\varepsilon}}
\def\engqks{{\mathfrak{K}}}
\def\Engk{{\mathcal{K}}}
\def\Engdk{{\mathcal{K}}}

\def\engks{{\mathfrak{K}}}

\def\Pq{\mathcal{P}}
\def\pq{\wp}
\def\pq{p}
\def\ps{p}

\def\V{V}
\def\Vol{{V\hspace{-0.3cm}^{_{\mbox{-}}} \hspace{0.13cm}}}

\def\Acuni{\Ac_{{\Ksub}^\dagsup}^{\dagsup}}
\def\unduni{\Ac_{{\Ksub}^\dagger}^{\dagsup}}
\def\Acauni{\Ac_{{\Ksub}^\ddagsup}^{\ddagsup}}
\def\Acunim{{\Ac_{{\Ksub}^\dagsup}^{\dagsup *}}}
\def\undunim{{\Ac_{{\Ksub}^\dagsup}^{\dagsup}}^*}
\def\Acaunim{{\Ac_{{\Ksub}^\ddagsup}^{\ddagsup *}}}
\def\pd{\partial}
\def\Ad{ {\mit \psi}}
\def\psim{ {\mit \psi}}
\def\Kd{K_d{}}
\def\Lam{{\mit \Lambda}}
\def\lam{\lambda}
\def\dagsup{{\mbox{\tiny${\dagger}$}}}
\def\ddagsup{{\mbox{\tiny${\ddagger}$}}}
\def\psimKdK{\psim_{\Ksub,\Kdsub}}
\def\w{\omega{}}
\def\wdlow{\omega_d }
\def\g{\gamma{}} 
\def\Phimc{{\mathcal C}}
\def\Psim{{\mit \Psi}}
\def\arm{{\rm a}}
\def\brm{{\rm b}}
\def\crm{{\rm c}}
\def\drm{{\rm d}}
\def\erm{{\rm e}}
\def\frm{{\rm f}}
\def\grm{{\rm g}}
\def\hrm{{\rm h}}
\def\lf{\left}
\def\rt{\right}
\def\Kdsub{{\mbox{\tiny${K_d}$}}}
\def\psimkd{\psim_{\kdsub}}
\def\psimKd{\psim_{\Kdsub}}
\def\hquad{ \ \ } 
\def\Taum{{\mit \Gamma}}
\def\dagsup{{\mbox{\tiny${\dagger}$}}}
\def\ddagsup{{\mbox{\tiny${\ddagger}$}}}
\def\muc{{\mu\hspace{-0.25cm}\mbox{-}\hspace{0.13cm}}}

%\tableofcontents

\begin{abstract}
Since its  discovery by Max Planck in 1900,  the  Planck constant $h$ has been demonstrated to be an universal constant, and its numerical value has  been  accurately determined based on  experiments. Up to the present however the physical origin of this fundamental constant  has not been well understood, and the numerical value of it has not been {\it ab initio}  predicted. $h$ is characteristic in two respects: 1) it is a universal constant with respect to all (quasi-) stationary dynamical processes of all matter particles and radiation fields, and 2) it has a specific numerical value. A theoretical inference of $h$, and a corresponding accounting  for the physical origin of $h$, therefore needs be achieved in  both respects. This paper presents a theoretical exploration in the first respect, a mathematical inference  of the universal constancy of $h$, based on the second law of thermodynamics, the principle of least action and the probability theory.

\end{abstract}

\section{Introduction}\label{Sec-b-intro}

Max Planck hypothesised energy quantum  in 1900\cite{Planck:1900} 
in order to resolve the then demonstrated large discrepancy between classical kinetic theory prediction and experimental data for black body radiation in the high frequency end, known as the "ultraviolet catastrophe". That  work laid the foundation of quantum mechanics. Through the formal quantum mechanics subsequently developed by E Schr\"odinger, W Heisenberg, P Dirac  and others in 1920s-30s along with a wide variety of experiments subsequently performed, it has become well established that  $h$ is a universal constant for all (quasi-) stationary processes of all matter particles and radiation fields.

At the scale $h$, a classically point matter particle turns to behave as a wave, $\Psim$. At this scale, the  motion of a particle $i$ of mass $\M_i$ and potential energy $\V$ is, in directly non-relativistic terms here, governed by the 
the Schr\"odinger equation, with $\hpbar_{k n} =\frac{\hp_{kn}}{2\pi}$, $\hp_{kn}=h$,
$$\displaylines{\refstepcounter{equation} \label{eq-Schra1}\label{eq-Schra}
\hfill
\imath \hpbar_{kn} \pd_t \Psim = H\Psim, \quad H= -\frac{\hpbar_{kn}^2}{2\M_i} \nabla^2 +\V.   
\hfill (\ref{eq-Schra1})
}$$
Electromagnetic radiation fields may be subjected to the same equation  (\ref{eq-Schra1}),  provided one regards, as is commonly done in quantum electrodynamics, (\ref{eq-Schra1}) as describing the motion of a charged harmonic oscillator emitting the radiation. In a similar way, (\ref{eq-Schra1}) presents also a governing equation for the generation of the total internal radiation fields, and hence the total energy and mass of a particle (see Sec. \ref{Sec-extn}). Our central concern in this paper is the fundamental Planck constant $h$. Equation  (\ref{eq-Schra1}) and other basic quantum-mechanical laws, on the other hand, present the context within which $h$ is meaningfully defined. It will suffice that we refer to  (\ref{eq-Schra1}) as representing the overall quantum-mechanical laws, since from  (\ref{eq-Schra1}) or the solutions to (\ref{eq-Schra1}), the basic dynamical  variables of a particle and their relations such as  the thermal and the total eigen energies and  momenta, and the Heisenberg uncertainty relations, among others, can all be derived.

It is well understood that the wave equation (\ref{eq-Schra1}) in essence describes the {\it stationary} state of a particle of a wave function $\Psim(x,t)$ and probability density  $\rho(x)=|\Psim(x,t)|^2$. This follows directly from the fact that every equation  (\ref{eq-Schra1}) is associated with a continuity equation for $\rho$, $\frac{\pd \rho}{\pd t}+\nabla (\rho \vel) =0$, where $\vel$ is the velocity of $\rho$. In a confined space, the dynamical variable solutions to (\ref{eq-Schra1}) for the particle are furthermore generally quantised, with respect to the Planck constant $h$ as prescribed in (\ref{eq-Schra1}). Evidently, however, a quantised stationary-state solution will result irrespective  of what the  $\hp_{k n}$ value is. $\hp_{k n}$ can in fact be an arbitrary real-valued parameter  for each specified energy level $n$ and (positional) degree of freedom $\j$. More generally, $\hp_{k n}$ can be dependent on the nature of the particle system and its environment, as Planck similarly had suggested in \citePlanck1900b. 
                          %\cite{Planck:1900}b (p.126) 

Today, $h$ is taken for granted to be a universal constant.  Its numerical value has  hitherto been determined by fitting such theoretical (or empirical) functions as the blackbody radiation spectrum,  and photo-electric work function, etc.,  to experimental data. These appear to have all resulted in $(\hp_\jn=)h =6.6260 ( ) \times 10^{-34}$  Js.                
          %(CODATA, 2002).
          %P J Mohr, B N Taylor, CODATA recommended values of the fundamental constants: 2002, Rev Mod Phys. vol 77 (2005)
          %*****************************************************************************
The physical origin of the $h$ in respect both to its universal constancy and its specific value, however, has hitherto remained an open question. There is no known logical reason why all $\hp_\jn$'s need be equal, in particular equal to $h$. 

The universal constancy of $h$ is also responsible for a basic concept in statistical mechanics, namely, in the phase space of, say, $N$ particles having $\f$ positional degrees of freedom each, there exists a smallest volume, $h^{\f N}$, which can be accessed by each microscopic 
          %(or quantum) 
state, or microstate, of the $N$-particle system. This concept was also originally introduced by Planck \citePlanck1900b, based on an hypothesis of absolute entropy and a presumption of the validity  of the postulate of  equal {\it a priori} probabilities, namely that all microstates  of an isolated thermodynamic system in equilibrium are equally accessible.  Similarly, there is hitherto no known logical reason why the involved postulate, which is mainly in question here,  holds except that based on it, statistical-mechanical solutions for thermodynamic systems have shown overall agreement  with experiment.

An understanding of the origin of the universal constancy of $h$ will shed light on one of the  key issues (another being the origin of energy quantisation) regarding the origin of quantum mechanics;  and this is also a  simultaneous step needed for an {\it ab initio} prediction of  the numerical value of $h$. In this paper, starting with a thermodynamic  system with its component particles described by the wave equation  (\ref{eq-Schra}) each, with $\hp_{\j n}$ assumed an arbitrary parameter, we shall infer the postulate of equal {\it a priori} probabilities (Secs. \ref{Sec-mic-ens}--\ref{Sec-extn}) and the universal constancy of  $\hp_{\j n}$, hence $h$ (Sec. \ref{sec-LeastAct}), based on the second law of thermodynamics and Maupertuis-Jacobi's principle of least action, severing as our first principles here, combined with the  probability theory. 

\section{Maximum entropy }\label{Sec-mic-ens}

We consider, until Sec. \ref{Sec-extn}, $N$ (identical) particles  in an enclosure of volume $\Vol$,   that are (a) isolated (from the environment), (b) weakly interacting only,  and (c) distinguishable. Here, (b) is said in the sense that the particles may exchange energies but present no correlations in respect to occupancy of states. The $N$-particle system is in (or approaching) thermodynamic equilibrium and is thus, owing to condition  (a),  of (or approaching) a fixed internal energy $U$. The particles are assumed to be each governed by the wave equations (\ref{eq-Schra}),  with  $\hp_{\j n}$  assumed arbitrary parameters. We shall regard the particles as intrinsically waves here;  at the large quantum number ($n$) limit these will reduce to the classical "point" particles.

Assume that the $N$  particles have each  $\f$ (positional) degrees of freedom, or dimensions. The quantum state of a particle $i$ ($i=1, 2, \ldots, N$) at time $t$ is thus completely specified by $D$ eigen wave functions $\Psim_{\j n_\jj}(x_{\j},t)'$s, with $\j= \f i-(\f -1),\f i-(\f-2),\ldots, \f i$.  In a confined space $[0, L_{\j }]$ 
(e.g. $L_\j=\Vol^{1/3}$ for a free particle in  $\Vol=L_\j^3$),   
 the eigen energy  $\Engvar_{\j n_\jj}$ of the particle $i$ is in general quantised,  
with $n_\jj =1,2,  \ldots,\nfrak_{\j o}$, where $n_\jj $ is  the principal quantum number\footnote{
           %**************
This holds also for  a particle described by a set of quantum numbers, e.g.  $n$, $l$, $m$, $s$ of a bound atomic electron.
Here, $l$ ($=0,1, \ldots, n-1$) and $m= \pm l$ reflect the geometric orientations of the total angular momentum ($n\hbar$) and total magnetic moment of the orbiting electron in magnetic field. $s$ describes the spin which is a permanent property of the electron  and, as of an any elementary particle, is never subjected to exchange.
%(although it may underlie the specific $h$ value). 
The $l,m, s$ each yield no independent or direct contributions to entropy, and thus need not be the concern of  this paper.} 
         %**************
 and $\nfrak_{\j o}$ is the total number of  possible  energy levels in $\j$th dimension.
There are a total  $\sum_{\j=Di-(D-1)}^{\f i} \nfrak_{\j o}$ number of distinct possible  states for particle $i$. 

 The microstate of the $N$-particles at a given time $t$ is accordingly completely specified by  $ \f \times N$  eigen wave functions of the $N$ individual particles.  Each $i$th particle at time $t$ statistically lies in a definite energy level $n_\jj$,  
of an energy  $ \Engvar_{\j n_\jj^{(i)}  }$, with $n$ suffixed for explicitness here by the particle index $(i)$. And the $N$-particles has a definite total internal energy   
$U=\sum_{i=1}^N\sum_{\j=Di-(D-1)}^{Di} \Engvar_{\j n_\jj^{(i)}}
                      %\sum_{\j=1}^{\f N} \Engvar_{\j n^{(i)}}
=\sum_{i=1}^N \Engvar_{i n^{(i)}}$.   Of the total $\nfrak_{\j o}$ levels for $i$th particle  in $\j$th dimension,  there can be $\nfrak_{\j}$  ($\le \nfrak_{\j o}$) levels that satisfy  condition (a),  given by   $n_\jj$'s lying  in a narrow range  $[\jj n_{a}, n_{\jj b}]$,  defining therefore an $N$-particle system with fixed $N,\Vol,U$. 
The corresponding possible microstates may be expressed by  the set of eigen wave functions
 $$\displaylines{\refstepcounter{equation} \label{eq-Kbb}
\hfill
\Bbb 
=
 \{\Psim_{1 n^{(1)}}, \ldots,  \Psim_{\f  n^{(1)}};  \ldots, \Psim_{\j n^{(i)}}, \ldots;   \Psim_{(\f N-(\f-1)) n^{(N)} }, \ldots, \Psim_{(\f N) n^{(N)}   } |   n_{ a}^{(i)} \le n^{(i)} \le n_{ b}^{(i)} \} 
              % \{\Psim_{1 n_1^{(1)}}, \ldots,  \Psim_{\f  n_\f^{(1)}};  \ldots, \Psim_{\j n_\j^{(i)}}, \ldots;   \Psim_{(\f N-(\f-1)) n_{(\f N-(\f-1))}^{(N)} }, \ldots, \Psim_{(\f N) n_{(\f N)}^{(N)}   } |   n_{\jj a} \le n_\jj\le n_{\jj b} \} 
                \qquad \quad 
\cr 
 \quad 
= \{\Psim_{1 \n}, \Psim_{2 \n}, \ldots, \Psim_{\j \n}, \ldots,  \Psim_{(\f N) \n}|\n=1,2, \ldots, \Nst \}
\hfill (\ref{eq-Kbb})
}$$ 
which make up a microcanonical ensemble. In the second of Eqs. (\ref{eq-Kbb}), each microstate  specified by the original set of  level  indexes $n^{(1)}, \ldots, n^{(i)}, \ldots, n^{(N)}$ is re-labelled, in an arbitrary order relative to the original,  by a single running index $\n$, $\n$ being an unique value for each distinct microstate. $\Nst$ denotes  the total number of possible microstates  satisfying the condition (a), and is formally given as 
$$\displaylines{\refstepcounter{equation} \label{eq-Nst}
\hfill
\Nst=\sum_{i=1}^N  \sum_{\j=Di-(D-1)}^{\f i} \nfrak_{\j }
= \sum_{i=1}^N  \Nst_i, 
%= \sum_{i=1}^N  (\nfrak \f)_{i}, 
\quad  
\Nst_{i} =\sum_{\j=Di-(D-1)}^{\f i} \nfrak_{\j }.
            %\nfrak_{i} =\sum_{\j=Di-(D-1)}^{\f i} \nfrak_{\j }.
\hfill (\ref{eq-Nst})
}$$ 
We assume here, as condition (d),  that $\Nst >> N$ so that the situation of more than one particle occupying the same state will be rare. The $\Nst_i$ or $\nfrak_\j$ wave functions of a particle $i$ span a Hilbert space in the state- or energy- representation,  where the state vector is
$\Psim_{\j, \ens}
=\sum_{n^{(i)}=1   }^{ \nfrak_\j } c_{\j n^{(i)} }  \Psim_{\j n^{(i)} }$, or
$\Psim_{i, \ens} 
=\prod_{\j =\f i-(\f-1)}^{\f i} \Psim_{\j, \ens}
               % = \sum_{n^{(i)}=1}^{\nfrak_\j}   \prod_{\j^{(i)}} \Psim_{\j \ens}  
 = \sum_{ n^{(i)}=1      }^{ \nfrak_{\j }} c_{i n^{(i)}} \Psim_{i n^{(i)}}$, 
$\Psim_{i n^{(i)}}=  \prod_{\j =\f i-(\f-1)}^{\f i}  \Psim_{\j, n^{(i)}} $.   
The $\Nst$ wave functions (\ref{eq-Kbb}) span a Hilbert space $\Hscr$ of the $N$  particles,   where the state vector is $\Psim_\ens=\prod_{i=1}^N \Psim_{i,\ens} 
= \sum_{\n=1}^{\Nst}  c_{\n} \Psim_\n$,           
$\Psim_\n=({\rm Sign}) \sqrt{r}  \Psim_{1\n}(x_1) \cdot \ldots \cdot \Psim_{(\subDN)\n} (x_\subDN) $
               %$\Psim_\n=\frac{({\rm Sign})}{\sqrt{N!}} \sqrt{r}  \Psim_{1\n}(x_1) \cdot \ldots \cdot \Psim_{(\subDN)\n} (x_\subDN) $
 being the total wave function of the $N$ independent and (for $r=1$)  distinguishable  particles here. The $c_\n$ etc. in the above are the amplitudes of states $\n$, etc. 

The microscopic states of the $N$-particles may be alternatively described in  an usual  $2\f N $--dimensional phase space $\Pscr$ spanned by  $\f N $ space coordinates 
%(with $\j=1, 2, \ldots, \f N$) 
$x_{1},  \ldots, x_{\f N}$  and $\f N $ momentum coordinates $\P_{1},  \ldots,  \P_{\f N}$ of the $N$ particles. A  $\n$th distinct microstate 
is in $\Pscr$ completely specified by a volume element 
$(\D x_{1\n},  \ldots,  \D x_{(\f N)\nu}$; 
$\D \P_{1\nu}, \ldots,  \D \P_{(\f N)\nu}) $ located about a fixed  point
$ (x_{1\nu},   \ldots,   x_{(\f N)\nu}$;
$\P_{1\nu}, \ldots,  \P_{(\f N)\nu})$, which are bounded between adjacent energy levels each as will be formally defined by (\ref{eq-DxDt})--(\ref{eq-engdif}); and this  has a one to one correspondence with a point $\Bbb_\n$,  
$= \{\Psim_{1 \n},   \ldots,  \Psim_{(\f N) \n     }  \}
$, in  $\Hscr$.
Accordingly, each $\n$th volume element  occupies a  volume 
  $$\displaylines{\refstepcounter{equation} \label{eq-bnux} \label{eq-B}
\hfill \b_\n= \prod_{\j=1}^{\f  N} \D x_{\j \n} \D \P_{\j \n} = \prod_{\j=1}^{\f  N}  \hp_{\j \n }, \quad   
\hp_{\j\n}=\D x_{\j \n} \D \P_{\j \n}, \quad  \n=1,2, \ldots, \Nst; 
\hfill (\ref{eq-bnux})
}$$
and all of the $\Nst$ volume elements occupy a total volume  $B=\sum_{\n=1}^\Nst \b_\n$ in $\Pscr$. By its definition (\ref{eq-bnux}), $\b_\n$ is the smallest  volume accessible to a state $\n$.  

At any instant of  time $t$  the $N$ individual particles statistically lie in their respective definite eigen states $ \Psim_{1 n^{(1)}}, \ldots, \Psim_{(\f N) n^{(N)} }  $ with the probabilities $ |c_{1 n^{(1)}}|^2, \ldots, |c_{(\f N) n^{(N)}}|^2$. 
And  the $N$-particles lie statistically in a definite microstate $ \Bbb_\n$, or volume  $\b_\n$ in $\Pscr$, with the probability $|c_\n|^2$. Over long time, 
as the result of  particle--particle interactions under condition (a),
the $N$-particles 
                      %(assuming Liouville) **No 
will statistically explore all the accessible possible  microstates.  To observe all of the $\Nst$ possible statistical states---providing also accessible--- at least once requires a measurement to be made over long time. 
To facilitate a measurement required only to be made at any one time,  
we may instead employ a  Gibbs ensemble consisting of $\Ng (\le \Nst)$ replicas of the original $N$-particle system, 
$$\displaylines{\refstepcounter{equation} \label{eq-Rbb}
\hfill
\Rbb= \{\Psim_{1\mu}, \Psim_{2\mu},    \ldots, \Psim_{\j \mu}, \ldots, \Psim_{(\f N) \mu}|\mu=1,2,\ldots, \Ng \}
\hfill (\ref{eq-Rbb})
}$$
which all have different microstates and yet identical macroscopic properties, the  fixed $N,\Vol,U$  here.

We now want to accommodate the $\Ng$ replicas of the $N$-particle system  in the phase space $\Pscr$, by   imagining  these as $\Ng$ objects  "thrown" into $\Pscr$, which will spontaneously attain their equilibrium positions after a relaxation time. Consistent with our presumption that  the $\hp_{\j\n}$'s are not necessarily all  equal (Sec. \ref{Sec-b-intro}), however, we do not {\it a priori} assume here, as the postulate of equal {\it a priori} probabilities instead does, that all the $\Nst$ possible microstates $\Bbb$ are equally accessible to the $N$-particles. Or equivalently, we do not assume that  all the $\Nst$ volume elements in $\Pscr$ can be readily uniformly occupied by  the $\Ng$ replicas, and this leads to  $\Ng\ne \Nst$ (or $\Ng < \Nst$). The actual accessibility or occupancy of $B$ by $\Rbb$, hence the actual correspondence between $\Bbb$ and $\Rbb$, is to be determined.

The degree of accessibility of a $\n$th state is directly proportional to the probability $\Pcal_\n$ for the $N$-particles to be found in the $\n$th state upon a measurement at time $t$, or equivalently, for a replica $\mu$ to be found in the $\n$th volume element in $\Pscr$. We assume that the $2\f N $--dimensional phase space $\Pscr$ in respect to a replica pertains to a geometric nature in the same sense as the position space in respect to, say, a dart in the dart game. Thus to attempt to accommodate or "throw" a replica $\mu$ in a volume element $\n$  in region $B$ in $\Pscr$ is like to attempt to throw a dart into  a small surface element of area $\sig(r,dr, d\theta)$  on a dartboard $\Dcal$. Based on dart game experiment (see e.g. \cite{Grinstead}) performed  in an effectively homogeneous space (assuming gravity field etc. is negligible), the larger the area $\sig$  is, the more likely the dart will hit $\sig$ on $\Dcal$. The probability $\Pcal_\sig$ for a successful hit at the target area  $\sig$ is thus $\Pcal_\sig \propto \sig $.
Making direct analogy, granted with a homogeneous phase space 
 in region $B$ where $U$ is everywhere the same, then the larger the volume $\b_\n$ of the $\nu$th volume element  is, the more likely a replica $\mu$ will "hit", or be accommodated  in, $\b_\n$.  That is, the probability $\Pcal_\n$ for a successful accommodation of a replica $\mu$ in $\b_\n$ is $ \propto \b_\n$. 
Multiplying $\b_\n$  by $1/B$ gives the normalised probability 
 $$\displaylines{\refstepcounter{equation} \label{eq-probn}
\hfill
\Pcal_\n = \frac{\b_\n}{B}, \quad \n =1,2, \ldots, \Nst.
\hfill (\ref{eq-probn})
}$$ 
Evidently, 
$\sum_{\n=1}^{\Nstsup} \Pcal_\n=(1/B)\sum_{\n=1}^{\Nstsup}\b_\n=1$.

Now according to the second law of thermodynamics, the disorderliness, hence  the entropy $S$, of an isolated system in (or approaching) thermodynamic  equilibrium  is (or approaches) maximum. Evidently, the disorderliness of the $N$-particles will be  maximum if the system explores as uniformly as possible all of the $\Nst$ possible microstates over  time. This  thus requires:
$$\displaylines{\refstepcounter{equation} \label{eq-statm} 
\quad
\mbox{
\begin{minipage}[b]{13. cm} 
{\small  All the $\Nst$ possible microstates are at any time $t$ simultaneously  equally accessible to  the $N$-particle system.}
\end{minipage}
}
\hfill (\ref{eq-statm})
}$$
  (\ref{eq-statm})  may be restated in terms of the $\Ng$ replicas and the phase space  as   
$$\displaylines{\refstepcounter{equation} \label{eq-statmb} \label{eq-Nssacc}
\quad
\mbox{
\begin{minipage}[b]{13.cm} {\small 
The $\Nst$ volume elements in $\Pscr$ are  simultaneously uniformly occupied 
by  $\Ng=\Nst$ replicas,  with one and only one replica occupying one  volume element at a time.}
\end{minipage}
}                   %\end{center}
 \hfill (\ref{eq-statmb})
}$$
The $\Nst$ simultaneously accessible states in (\ref{eq-statm}), or 
                       %equivalently but less abstractly 
the simultaneous occupancies of the $\Nst$ possible states 
                %by the $\Ng $ replicas of 
stated less abstractly in (\ref{eq-statmb}), represent  $\Nst$ simultaneous events that  are, due to condition (b), independent  with one another. The independence here removes the possibility of correlated occupancies of $B$ by the replicas due to particle interactions. The condition (d) restrains two or more replicas $\mu, \mu'$ from occupying the same state $\n$ at the same time. And the $\Nst$ value is specified by the number of eigen solutions of  (\ref{eq-Schra}) for the $N$ particles under the fixed  $N,\Vol, U$ condition. On the other hand, it remains possible that some regions in region $B$ in $\Pscr$ are densely populated and some other regions  rarely as a result of the geometric property of the phase space, so that effectively $\Ng < \Nst$; this situation dissatisfies (\ref{eq-statmb}) or (\ref{eq-statm}). If such situation is to be avoided, the phase space must possess certain quality  as may be identified as follows: 
The probability for the occurrence of the $\Nst$ simultaneous independent 
events  (\ref{eq-statm}) or  (\ref{eq-statmb})
 is according to probability theory given by
$$\displaylines{\refstepcounter{equation} \label{eq-Pcal1}
\hfill
\Pcal_{\ens}
=\prod^{\Nstsup}_{\n=1} \Pcal_\n
=\prod^{\Nstsup}_{\n=1} \frac{\b_\n}{B}
= \frac{\b_\Pisub}{B^\Nstsup},
 \quad 
\b_\Pisub
= \prod_{\n=1}^\Nstsup \b_\n. 
 \hfill (\ref{eq-Pcal1})
}$$ 
(\ref{eq-statm}) or (\ref{eq-statmb}) will be maximally fulfilled if $\Pcal_{\ens}$ is  maximum; and 
%whereas 
$\Pcal_{\ens}$ is a maximum ($\Pcal_{\ens.max}$) if 
$$\displaylines{\refstepcounter{equation} \label{eq-Pcalx1b2}
\hfill
\delta \Pcal_\ens =0, \quad \delta ^2 \Pcal_\ens <0.
\hfill (\ref{eq-Pcalx1b2})
}$$

To (universally) determine $\Pcal_{\ens.max}$ by solving (\ref{eq-Pcalx1b2}) using calculus will inevitably be unfeasible, since the volume elements in $\Pscr$ can be in arbitrary shapes owing to the diverse shapes and dynamics of physical systems and   $\Pcal_\ens$ is therefore not generally expressible into a universal analytical function. A maximum $\Pcal_\ens$ solution however can be obtained readily by  means of algebraic method  as follows. The $\Pcal_\ens$ given by  (\ref{eq-Pcal1}) is equivalent to the geometric mean of $\Pcal_\n$'s
 $$\displaylines{\refstepcounter{equation} \label{eq-Nstbar}  %\label{eq-Nst2}
\hfill
 \overline{\Pcal_\n}
= (\prod^{\Nstsup}_{\n=1} \Pcal_\n)^{\frac{1}{\Nstsup}}
=\frac{             (\prod^{\Nstsup}_{\n=1}\b_\n)
^{   1/\Nst } }{B}, \hfill (\ref{eq-Nstbar})
}$$
raised to the power of  $\Nst$, i.e.  $\Pcal_\ens=( \overline{\Pcal_\n})^\Nstsup $.
On the other hand, the arithmetic average is 
$$\displaylines{\refstepcounter{equation} \label{eq-Nst1}
\hfill
\la \Pcal_\n\ra= \frac{1}{\Nsts}\sum_{\n=1}^{\Nstsup}\Pcal_\n=  \frac{1}{\Nsts} \frac{\sum_{\n=1}^\Nst\b_\n }{B} = \frac{1}{\Nsts}; \quad {\rm thus } \ \Nst=
                     %\Nst_{i.a.m} \equiv 
\frac{1}{\la\Pcal_\n\ra}. 
               \hfill (\ref{eq-Nst1})
}$$ 
It follows from the theorem of inequality of the arithmetic-geometric means that 
$\overline{\Pcal_\n}  < \la\Pcal_\n\ra  = (\frac{1}{\Nst})$, and hence  
$\Pcal_\ens = \overline{\Pcal_\n}^{\Nstsup} < \langle\Pcal_\n\rangle^{\Nstsup} (= (\frac{1}{\Nst})^{\Nstsup})$, if the $\Pcal_\n$'s, being nonnegative real functions,  are not all equal. 
   And, 
$$\displaylines{
\refstepcounter{equation} \label{eq-Pcalx1}
\refstepcounter{equation} \label{eq-Pcalx2}
\hfill
 \overline{\Pcal_\n}{}_{.max}  =\langle\Pcal_\n\rangle=\frac{1}{\Nsts}, \quad 
  \Pcal_{\ens.max} (= {\overline{\Pcal_\n}{}_{.max} }^{\Nstsup})=  \la\Pcal_\n\ra^{\Nstsup}= \left(\frac{1}{\Nsts}\right)^{\Nstsup},
 \hfill (\ref{eq-Pcalx2})
}$$ 
i.e. $ \overline{\Pcal_\n} $ and $ \Pcal_{\ens}$ are  maxima each,  
if and only if the nonnegative real functions $\Pcal_\n$'s are all equal, $\Pcal_1=\Pcal_2=\ldots=\Pcal_{\Nst}$. Or equivalently,  (\ref{eq-Pcalx2}) holds if and only if, on substituting  (\ref{eq-probn}) for $\Pcal_\n$, all $\b_\n$'s are equal to one another and hence to a common value denoted by $\b_0$, 
$$\displaylines{\refstepcounter{equation} \label{eq-Pcalx1b}
\hfill
\b_1=\b_2=\ldots=\b_\Nst =\b_0.
\hfill (\ref{eq-Pcalx1b})
}$$
              %Here $\b_0$ designates  the common value of the $\b_\n$'s. 
With (\ref{eq-Pcalx1b}), (\ref{eq-B}b) becomes $B=\sum_{\n=1}^{\Nst}\b_\n=\Nst \b_0$
and (\ref{eq-probn}) becomes  
 $$\displaylines{\refstepcounter{equation} \label{eq-probnp}
\hfill 
 \Pcal_\n =\frac{\b_0}{\Nst \b_0} =\frac{1}{\Nst}, \quad \n =1,2, \ldots, \Nst. 
\hfill (\ref{eq-probnp})
}$$
(\ref{eq-probnp}) states that, for the $N$-particles under conditions (a)--(d), all the $\Nst$ possible microstates are equally accessible, which is just the statement of the  fundamental postulate of equal {\it a priori} probabilities. The entropy of the $N$-
particles follows to be a maximum, given as  
$$\displaylines{\refstepcounter{equation} \label{eq-S}
\hfill
S=-k_\Bsub \sum_{\n=1}^{\Nst} \Pcal_\n \ln \Pcal_\n = k_\Bsub \ln \Nst \sum_{\n=1}^{\Nst} \frac{1}{\Nst}= k_\Bsub \ln \Nst. 
\hfill (\ref{eq-S})
}$$
The maximum probability $ \Pcal_\n$ and entropy  $S$ of  (\ref{eq-probnp})--(\ref{eq-S}) are achieved each in a statistical sense, as are the statements (\ref{eq-Pcal1}) and (\ref{eq-Pcalx1b2}), and hence allow for statistical fluctuations.

The formal steps from (\ref{eq-Pcal1}) to (\ref{eq-Pcalx2}) in the above have yielded  a  few significant equations (\ref{eq-Nstbar}), (\ref{eq-Nst1}) and (\ref{eq-Pcalx2}).
Without these we could have as well obtained (\ref{eq-Pcalx1b}) and (\ref{eq-probnp}) by continuing the geometrical argument as made for (\ref{eq-probn}). Namely, because $\Pcal_\n \propto \b_\n $,  all $\Pcal_\n$'s are thus equal, whence  satisfying   (\ref{eq-statm}) or (\ref{eq-statmb}), if all $\b_\n$'s are equal. 

\section{Global system}\label{Sec-extn}
For defining $\Nst$ and $\b_\n$, and for arriving at the basic conclusions in Sec. \ref{Sec-mic-ens},  Eq. (\ref{eq-probnp}) in particular, the $N $ particles need not be all identical. The $N$ particles may be of different  species,  chemical compositions and/or energy forms (see below), and they may be of different interaction and  motion schemes. These may for example include $N_a$ atoms, $N_e$ electrons (free or bound), $N_{pht}$ photons (electromagnetic radiation energy quanta), $N_{phn}$ phonons (sound energy quantum), and so forth. Although, 
the total quantities of the mixed particle system need  be given by the sums  over all particle species, etc., e.g. $N=\sum_{\a=1}^3 N_\a$, $\Nst= \sum_\a N_\a \f_\a \nfrak_\a$. And  mixed particles will typically involve different  degrees of freedom (DOF's).  If two different  DOF's represent the same energy,  such as the energy levels of a harmonic oscillator and the quanta of its radiation fields, evidently only one should be considered for entropy. If a DOF is  internal, such as the orbiting motion of an atomic  electron, this DOF should not be (directly) considered for entropy.

In usual applications, Eq.  (\ref{eq-Schra1}) describes the kinetic motion of a matter particle such as  an electron in an applied potential field $\V$, or indirectly the  thermal radiation fields. The  $E_{in}$, $P_{in}$ etc. are the dynamical variables associated with such motions. Alternatively,  (\ref{eq-Schra1})  also describes  the motion of  a charge, of a dynamical mass $\Mcal_{qi}$ (in the place of $M_i$),  in a (quadratic) vacuum potential field across a distance  $b_\v \sim 10^{-18}$ m, which generates in terms of the IED model \cite{jxzjied} the total, internal motion of a simple matter particle such as an electron, proton (see a systematic treatment in \cite{jxzjied}c).  $(\Engvar_{in}\rightarrow) \Eng_{q_in}=n\hbar \W_i=nM_ic^2 $ is then the  total energy and $P_{in}$ the total linear momentum of the particle, with $\W_i=\frac{M_ic^2}{\hbar}$, and $M_i$ (with $n=1$) the mass of the resulting stable IED particle. The basic conclusions of Sec. \ref{Sec-mic-ens}  apply evidently directly to  this latter system ---a particle of an alternative (total as contrasted to thermal) energy form.

Equation (\ref{eq-probnp}) has been otherwise arrived at for a system of $N$ particles
under the conditions (a)--(d) that are also  the conditions for the direct application of the postulate of equal {\it a priori} probabilities in statistical mechanics. So (\ref{eq-probnp}) may be extended to other (important) variant   systems following  the same well-established "correction" procedures (see e.g. \cite{Hill1960}) in statistical mechanics, as are briefly outlined in (i)--(iv) below. (i) Two important variations from  the "isolated system" of cond.  (a) are the "closed isothermal system", whose $U$ may fluctuate and $N,\Vol,T$ are fixed, and "open isothermal system", whose $U,N$ may fluctuate and $\Vol,T,\muc$  (where $\muc$ is the chemical potential) are fixed. The $U$ and $U,N$ fluctuations yield inhomogeneities in $\Pscr$, their  $\Pcal_\n$'s are now given by    the canonical (or Boltzmann) and grand canonical distributions
$$
\displaylines{\refstepcounter{equation} \label{eq-Q1}
\hfill 
\Pcal(U)                      %=\frac{\sum_\n^{\Nst(U)} e^{-U_\n/k_\Bsub  T}}{ Q}
 = \Nst(U) e^{-\frac{U}{k_\Bsub  T}}/Q(T,\Vol,N), 
\quad  \Pcal(U,N) =\Nst(U,N)e^{-\frac{U-\mu N}{k_\Bsub T}} /Z(\Vol,T,\mu), 
\hfill(\ref{eq-Q1})
}$$
where
$Q
                   %=\sum_{U'} \sum_{\n=1}^{\Nst(U')} e^{-\frac{U'_\n}{k_\Bsub T}}
=r \sum_{U'} \Nst(U')e^{-\frac{U'}{k_\Bsub T}}$,  and  
$ 
Z= r\sum_{U',N'}\Nst(U',N')e^{-\frac{U'-\mu N'}{k_\Bsub T}} 
$,
with $r=1$ and $1/N!$ for $N$ distinguishable and indistinguishable particles. 
                        %ch Reif, p131
Separately, the  $N$-particle system may change energy, by $\D U$, by interacting with  another system by means of thermal interaction under fixed  external  parameters, or  by adiabatic change of  external parameter(s), or by a combination of the two. 
In any of the cases,  $\Nst$ is unchanged. 
(ii)  The $N$ particles may be correlated with one another, hence a deviation from cond.  (b). The usual method to extend  beyond (b) is  virial expansion of the equation of state.
               % (\cite{Hill1960}, chap 15).  
(iii) The $N$ particles may be indistinguishable with one another, hence a deviation from cond. (c). This is especially the feature with many quantum particles confined in a common small geometry,  where their wave functions mutually overlap to a high degree. The standard correction to (c)  is, as is similarly done in (\ref{eq-Q1}), by means of dividing out $N!$ number of indistinguishable states from $\Nst$, giving $\Nst_{I.D}=\frac{\Nst}{N!}$.  
(iv)
$T$ may be very low such that 
$\Nst \sim N$ and  the situation of more than one particle occupying the same states becomes frequent,  hence a deviation from cond. (d). The standard approach to such system is  the replacement of  the Boltzmann statistics by  quantum statistics (Fermi-Dirac or Bose-Einstein statistics) where restrictions are  made to the number of accessible states by applying symmetry properties of the wave functions of particles. 
Of  the systems of (i)--(iv), the $b_\n$ and $  \Nst$ values  are unchanged.   
Only the  extents or forms of the contributions of their  $\Nst$ to $\Pcal_\v$ may be  modified and thus corrected for. The corrections may be appreciated as being to scale the systems back  to obeying conditions (a)--(d). A recognition of this feature permits  one to generalise the approach of "scaling back" also to other possible variant systems not yet considered. 

The diverse  forms of particle systems of this section and Sec. \ref{Sec-mic-ens} together resemble a major  important part of the "global system" of the physical world in which  energy transformations and exchanges in units of  integer times an energy quantum $\hbar  \times $ angular-frequency (e.g. the $\W_i$ earlier)  have been observed. To this global system the inference of  the postulate of equal {\it a priori} probabilities in Sec. \ref{Sec-mic-ens}, Eq. (\ref{eq-probnp}), is  valid in the fashion as case by case discussed above.

\section{Least action 
}
\label{Sec-Heisenberg}
\label{subsec1}
\label{sec-LeastAct}

Insofar as all the $\b_\n$'s are ensured equal (to $\b_0$) and thus $S(U)$  is maximum for a prescribed  $\Nst$ value, it is irrelevant that what the $\hp_{\j\n}$ values, and hence the $\D x_{\j\n}$ and $ \D \P_{ \j \n}$ values according to  (\ref{eq-bnux}), are unless otherwise constrained.  In the following  we shall not assume any pre-established knowledge of the $\hp_{\j\n}$'s (esp. their being equal to $h$), but shall investigate the characteristics of the $\hp_{\j\n}$'s,  or the  $\hp_{\j n}$'s below, regarding  their constancy based on Maupertuis-Jacobi's least action principle

Of the $N$ particles of Sec. \ref{Sec-mic-ens}, we focus now on an individual particle ($i$)   in motion at a component velocity $\vel_k$  in $\j$th dimension confined in a space interval $[0,L_\j]$. The particle is assumed (quasi-) stationary and described by an eigen wave function $\Psim_{\j n}$ and  probability density $|\Psim_{\j n}(x_\j,t) |^2$ governed by Eq.  (\ref{eq-Schra}). The so-described particle is generally extensive across $[0,L_\j]$ at any time $t$, and oscillating over all time at any location $x_\j$ in $[0,L_\j]$. Its dynamical variables, such as linear momentum $\P'_{\j n}(x_\j,t)=\M_{i }\vel'_{\j n}$, kinetic energy $\Engk'_{\j n}(x_\j,t)=\frac{1}{2}\P'_\jn(x_\j,t)\vel'_\jn(x_\j,t)$, 
potential energy $\V'_{\j n} (x_\j(t))$, and  Hamiltonian  $\Engvar'_{\j n}(x_\j,t)=\Engk'_{\j n}(x_\j,t)+\V'_{\j n} (x_\j(t)) $, are accordingly distributed functions in $[0,L_\j]$. 

For this undulatory extensive stationary particle, there exists in general a characteristic space interval $\D X_{\j n}$, such that only  over $\D X_{\j n}$, or $[0,\D X_{\j n}]$,  the particle can  be wholly and meaningfully defined, or in fact exists in stationary state when condition for complete constructive/destructive interference presents. For a variety of monochromatic periodic particle processes, $\D X_{\j n}=n\D X_{j 1}$; and the minimum of  $\D X_{\j n}$, $\D X_{\j 1}$, is given by  the distance of one-full cycle of the periodic process in space, which  is e.g. one wavelength of a plane de Broglie wave (this  however does not apply  to a plane electromagnetic wave of one energy quantum, or a photon, which is not directly governed by Eq \ref{eq-Schra1}), or $4\times $ the amplitude of a harmonic oscillation. There thus exists a corresponding characteristic time interval $\D T_{\jn } = \D X_{\jn } /\la\vel_\jn'\ra $  needed for the particle to traverse $[0, \D X_{\jn }]$. And in the case of $\D X_{\j n}=n\D X_{j 1}$, $\D T_{\jn } =n \D T_{\j 1}$; 
the minimum of $\D T_{\jn } $,   $\D T_{\j 1}$, is equal to one full period of time.  
              %An integral  $n$ multiples  of the  $\D X_{\j 1}$, $\D X_{\j n}= n\D X_{\j 1}$, at a given $t$, or similarly $\D T_{\j n}=n \D T_{\j 1}$,  will also specify the dynamical system  wholly meaningfully, although  this carries now $n$ times the minimal energy; or it carries the minimal meaningful  energy of $n$th level.  
$\D X_{\jn}$ for a particle at a given energy level $n$ must in turn be accommodated in an external environment, such as between enclosure "walls" at $x_\j=0$ and $L_\j$. Hence $\D X_{\jn}=\D X_{\jn}(L_\j)$.  For example, for a free particle of a wave function $\Psim_{n}(x,t)=C e^{ i(\frac{2\pi }{\Lam_{dn}} x - \frac{\Engvar_n}{\hpbar_{ n}}  t)}$ and  wavelength $\Lam_{dn}$ confined in a one-dimensional box $[0,L]$ (with the subscripts $\j$ omitted here), we find  $\D X_n =n  \D X_1$, $\D X_1=\Lam_{d1} $. But $\Lam_{dn} =\frac{2L}{n}$ as   given by the boundary conditions $\Psim(0)=\Psim(L)$. Hence  $\D X_n (= n \Lam_{dn})  = 2L$.

We define for the particle a Maupertuis-Jacobi's action integral\footnote{
$\Ia_{\j n}$ may be re-written as $\Ia_{\j n} 
= \int [\Engk'_\jn -V'_\jn + \Engk'_\jn + V_\jn']dt 
=\int [\Lcal'_\jn +\Engvar'_\jn]dt$, 
where $\Lcal'_\jn = \Engk'_\jn -V'_\jn$ is the Lagrangian function.   $\int \Lcal'_\jn dt =S_\jn$ defines a distinct action integral which is widely in use today and which, as may be  easily seen, is yet unsuited for the present problem.
}
  in $[0,\D X_{\jn}] $ as,
$$\displaylines{
\refstepcounter{equation} \label{eq-Ijacobi-Int}
\hfill
\Ia_\jn= \int_{0}^{\D X_{\jn}} \P'_{\jn}(x_\j,t)dx_\j
           %, \quad {\rm or} \ \Ia_\jn
=\int_{0}^{\D X_{\jn }} \frac{2\Engk'_{\jn} (x_\j,t) }{\vel'_\jn(x_\j,t)} dx_\j
= \int_{0}^{\D T_{\jn  }} 2\Engk'_{\jn} (x_{\j},t)d t, 
                              \hfill(\ref{eq-Ijacobi-Int})
\cr\mbox{where $dx_\j/\vel'_\jn =dt$. (\ref{eq-Ijacobi-Int}) may be re-written as} \hfill
\cr
\refstepcounter{equation} \label{eq-Ijacobi-Int2}
\hfill
 \Ia_\jn= \P_\jn \D X_{\jn}, 
\quad 
\Ia_\jn =(2 \Engk_{\jn}) \D T_{\jn},
 \hfill (\ref{eq-Ijacobi-Int2})
}$$
where $\P_\jn, \Engk_\jn$ are the expectation values of  $\P'_\jn, \Engk'_\jn$ or of the corresponding operators $\hat{\P}_\jn, \hat{\Engk}_\jn$ acting on $\Psim_{\j n}$ (with  $\Psim_\jn$ assumed normalised in $[0, \D X_\jn]$):
$$\displaylines{
\refstepcounter{equation} \label{eq-expcs}
\hfill
\P_{\jn}=\frac{1}{\D X_{\jn}}\int_0^{\D X_{\jn }} \P'_\jn(x_\j,t) dx_\j 
            =\int_0^{\D X_{\jn}} \Psim_\jn^*(x_\j,t) \hat{P}_\jn \Psim_{\j n}(x_\j,t)  dx_\j, \hfill \cr
\hfill 
 \Engk_{\jn}
=\frac{1}{\D T_{\jn}} \int_0^{\D T_{\jn}} \Engk'_{\jn}(x_{\j},t) dt 
= \int_0^{\D T_{\jn}}   \Psim^*_\jn(x_\j,t)  \hat{\Engk}_\jn  \Psim_\jn(x_\j,t)  d t;  \  
         %= \int_0^{\D T_{\jn}}   [\Psim^*_\jn(x_\j,t)  \hat{\Engk}_\jn  \Psim_\jn(x_\j,t)]_{x_k=0}  d t;  \  
\Engvar_{\jn}=\Engk_{\jn}+\V_\jn.
\hfill (\ref{eq-expcs})
}$$
           %$\Psim_\jn$ is  here assumed normalised in $[0, \D X_\jn]$.  
             % For  a stationary state, $ \Engvar_{\jn} $ is independent of time. 
($\P_{\jn}$ and 
$ \Engk_{\jn}$ of the typical stationary particle systems in applications, and   $\Engvar_{\jn}$ of any such systems, are in fact independent of $x_\j,t$.)
We further define the difference (Maupertuis-Jacobi's)  action integral 
as the absolute difference between $ \Ia_\jn$'s of two adjacent levels $n+1 $ and $n$:
$$\displaylines{
\refstepcounter{equation} \label{eq-Ijacobi-I}
\hfill 
\hp_\jn= |\Ia_{\jn+1}-\Ia_{\jn}|; \quad  {\rm or} \
\hp_\jn=  \D \P_\jn \D x_\jn
\equiv  \int_0^{     \D x_\jn}  \D \P'_\jn(x_\j,t)  dx_\j, \hfill\cr
\hfill 
 \hp_\jn
= (2\D \Engk_{\jn}) \D t_{\jn}  
\equiv \int_0^{\D t_{\jn}} 2\D \Engk'_{\jn} (x_\j,t)d t 
\equiv   \D  \Engvar_\jn  \D t_{\jn},  
                %= \int_0^{\D t_{\jn}} \D \Engvar'_{\jn}d t, 
\hfill 
(\ref{eq-Ijacobi-I})
}$$
where  $\D x_\jn$ and $\D t_\jn$ are  the mean characteristic space and time intervals defined by
$$\displaylines{
\refstepcounter{equation} \label{eq-DxDt}
\hfill
\D x_\jn= \frac{|\P_{\jn+1} \D X_{\jn+1  } -\P_\jn \D X_{\j n}|        }{ \D \P_{\jn} }, 
\quad 
\D t_\jn 
=\frac{|\Engk_{\jn+1} \D T_{\jn+1 } - \Engk_{\jn} \D T_{\jn } |    }{\D \Engk_{\jn} };
 \hfill (\ref{eq-DxDt})
\cr
{\rm and}\hfill \cr 
\refstepcounter{equation} \label{eq-engdif} \hfill
  \D \P_{\j n} = |\P_{\j (n+1)}- \P_{\j n}|, \quad
\D \Engk_{\jn}=|\Engk_{\jn+1}-\Engk_{\jn}|,
\quad \D \Engvar_{\j n}=|\Engvar_{\j( n+1)}-\Engvar_{\j n}|; \qquad
             %==\D  \Engk_\jn+ \D \V_\jn = 2 \D \Engk_\jn, 
\hfill (\ref{eq-engdif})
\cr
\refstepcounter{equation} \label{eq-engk}
\hfill \D \Engk_{\jn}= \D \V_{\jn}, \quad 
{\rm thus} \  
\quad \D \Engvar_{\jn}=\D  \Engk_\jn+ \D \V_\jn = 2 \D \Engk_\jn.
\hfill
(\ref{eq-engk})
}$$
$  \D \P'_\jn $ and $\D \Engk'_{\jn}$  are similarly defined. As it may be checked against  the typical quantum systems in applications, using $\D x_\jn, \D t_\jn $ defined in  (\ref{eq-DxDt}) as combined with the eigen solutions for $P_n, \Engk_n$ will indeed yield the exact forms of Heisenberg relations.
For a spatially confined  particle, the  $\D \P_{\j n} $, $\D X_\jn$,  $ \hp_\jn$, etc, are quantised as are the $P_\jn, \Engk_\jn$ and $ \Engvar_\jn$ (Sec. \ref{Sec-mic-ens}) and $\Ia_\jn$.  Moreover, the  $ \hp_\jn$, $\D \P_{\j n} $, $\D X_\jn$, etc are in the above each  positively defined since their magnitudes only, not  their signs, 
 will be relevant. The mechanical energy difference $\D \Engvar_{\jn}$  
corresponds to   an energy quantum of {\it harmonic} radiation field, or a photon, emitted upon  transition from $n+1$ to $n$, for which the relations (\ref{eq-engk}) are always valid.

It is by the requirement of the least action principle  that the $\Ia_\jn$ and accordingly $\hp_\jn$ need be minimum each and hence satisfy the conditions 
$$\displaylines{
\hfill (a): \quad  \delta  \Ia_\jn=0, \quad \delta^2 \Ia_\jn>0; 
\qquad  
\refstepcounter{equation} \label{eq-Ijacobi}
(b): \quad 
\delta \hp_\jn=\delta \Ia_{\j n+1}-\delta \Ia_{\jn} 
=0, \quad \delta^2 \hp_\jn >0.
\hfill (\ref{eq-Ijacobi})
}$$
   Since $\D x_\jn$, $ \D t_\jn$, $\D \P_{\j n}$ and $\D \Engvar_{\j n}$ are finite for the  
   $\D X_{\j n}$ and $\D T_{\j n}$ being finite---intrinsically for an intrinsically extensive IED particle\cite{jxzjied}, (\ref{eq-Ijacobi}) must have nontrivial, finite valued solutions. 
Since $\sum_{n'=0}^{n }\hp_{\j n'} =\Ia_\jn$, to achieve a minimum $\Ia_\jn$ solution for (\ref{eq-Ijacobi}a) it suffices, ultimately for the final solution form, that each $\hp_\jn$  is minimised according to (\ref{eq-Ijacobi}b).

For $N$ particles in the generalised sense of the global system of Sec. \ref{Sec-extn},  least action  should ultimately be satisfied by all of the $\hp_{\j n^{(i)}}$'s of all  energy levels of all $N$ particles, and  hence by the sum  of  $\hp_{\j\n}$'s over all $\n$ values:
$$\displaylines{
\hfill
\Hn = \sum_{\j=1}^{\f N} \hp_{\j \n}=\sum_{\j=1}^{\f N}
\D \P_{\j\n} \D x_{\j\n}
=\sum_{\j=1}^{\f N}2\D  \Engk_{\j\n} \D t_{\j\n}; \quad 
              \refstepcounter{equation} \label{eq-leastact2}
\delta \Hn =\delta \sum_{\j=1}^{\f N} \hp_{\j\n}=0, 
\quad \delta^2 \Hn >0.  
\hfill (\ref{eq-leastact2})
}$$
Similarly as for $\Pcal_\n$ earlier, a general differentiable function for $\Hn$ for all particles in the global system, which in general may be of arbitrary dynamics
and enclosure geometries, and thus a general solution for (\ref{eq-leastact2}b) given in terms of  calculus,  will be absent.  However, a  minimum  $\Hn$ can be readily obtained using  algebraic method as follows. 

In  (\ref{eq-bnux}) we have already written down a product  equation of  all $\hp_{\j\n}$'s, with the $\hp_{\j\n}$'s positively defined. With (\ref{eq-Pcalx1b}),  (\ref{eq-bnux}) is rewritten as
$$\displaylines{\refstepcounter{equation} \label{eq-h0}
\hfill
\b_\n=\prod_{\j=1}^{\f N} \hp_{\j\n}=\b_0, \quad \n=0,1,\ldots,  \Nst.                 %\nfrak 
\hfill (\ref{eq-h0})
}$$
For a chosen global system of a total fixed $N$ number of particles in or approaching equilibrium with fixed macroscopic properties, the product of  the $\f N $ parameters $\hp_{\j\n}$'s, $\b_0$,  given in (\ref{eq-h0}) is a fixed value. Then, according to the theorem (a corollary of the theorem of  inequality of the arithmetic-geometric means) for finding extrema,
                 % (see e.g. \cite{Niven}), 
 the sum $\Hn$ of the  $\f N $ positive $\hp_{\j\n}$'s as given by  (\ref{eq-leastact2}a) is a minimum if and only if all the positive $\hp_{\j\n}$'s are equal to one another, and hence to a common value denoted by $\hp_0$,  
$$\displaylines{\refstepcounter{equation} \label{eq-ajmu-a0}          
   \label{eq-ajmu-a0-2}    
\hfill 
\hp_{1\n}=\hp_{2\n}= \ldots =\hp_{\mbox{{\tiny{$(\f N)$}}} \n} =\hp_{0},  
\quad \n=1,2, \ldots, \Nst; 
\quad 
 \Hn{}_{.min} =\sum_\j^{\f N} \hp_0=\f N  \hp_0.   \hfill (\ref{eq-ajmu-a0-2})
}$$
% $\hp_0$ denotes the common value of $\hp_{\j \n}$'s. 
Substituting  (\ref{eq-Ijacobi-I}b)--(c)  in (\ref{eq-ajmu-a0-2}a)--(b) gives   
$$\displaylines{\refstepcounter{equation} \label{eq-ajmu-a0p}          
   \label{eq-ajmu-a0-2p}    
\hfill
\D P_{\j 1} \D x_{\j 1}=\ldots=\D \P_{\j\n} \D x_{\j\n}
=\D  \Engvar_{\j 1} \D t_{\j 1}= \ldots =\D  \Engvar_{\j \n} \D t_{\j \n }=\hp_0,  \quad \n=1,2, \ldots, \Nst; 
\hfill
\cr  
\hfill 
\sum_{\j=1}^{\f N}
\D \P_{\j\n} \D x_{\j\n}
=\sum_{\j=1}^{\f N}2\D  \Engk_{\j\n} \D t_{\j\n}=\f N \hp_0.
\hfill  (\ref{eq-ajmu-a0-2p})
}$$
Substituting (\ref{eq-ajmu-a0-2}) in (\ref{eq-Ijacobi-I}a)  gives  (for $\Ia_{n+1}>\Ia_n$) $\Ia_{\jn+1}=\Ia_\jn+\hp_0$. Calculating $\Ia_{\jn+1} $ 
successively, we obtain $\Ia_{\j 1}=\Ia_{\j 0}+\hp_0$, $ \Ia_{\j 2}=\Ia_{\j 1}+\hp_0=\Ia_{\j 0}+2\hp_0$, and so on; and 
$$\displaylines{\refstepcounter{equation} \label{eq-An}
\hfill
 \Ia_{\j n}(=\P_{\j n} \D X_{\j n}= 2K_{\j n} \D T_{\j n})=\Ia_{\j 0}+n\hp_0 
\hfill (\ref{eq-An})
}$$ 
Conversely, if the sum $\Hn $ of the  positive  $\hp_{\j\n}$'s is set to a fixed value, concretely the minimum $\Hn{}_{.min}$, then according to the foregoing theorem stated in reverse order, the product  of the positive $\hp_{\j\n}$'s, $\b_0(=\Pi_{\j=1}^{\f N} \hp_{\j\n})$, is a maximum if and only if all the positive $\hp_{\j\n}$'s are equal, as given in (\ref{eq-ajmu-a0}). With (\ref{eq-ajmu-a0}) and (\ref{eq-ajmu-a0-2p}), (\ref{eq-h0}) is written as 
$$\displaylines{\refstepcounter{equation} \label{eq-bnuxpp}
\hfill 
\b_0=\prod_{\j=1}^{\f N} \hp_{\j\n}
=\prod_{\j=1}^{\f N} \D \P_{\j\n} \D x_{\j\n}=\prod_{\j=1}^{\f N} 2\D  \Engk_{\j\n} \D t_{\j\n}
=\hp{}_0^{\f N}=\b_{\n.max}. 
 \hfill (\ref{eq-bnuxpp})
}$$

Except for being required as general and as inclusive as possible, 
the global system may be chosen differently at different occasions or  times, whose $\b_0$'s will thus be different.  
As long as the different global systems 
share at least one common particle---a trivial condition which can be readily satisfied in both theory and practice, however, the same universal $\hp_0$ must be preserved for all choices of the global systems at all times. $\hp_0$ thus is universal irrespective of the choice of the global system and time. 
And the foregoing inference of the universal constancy of $\hp_0$ is valid to the extend that the second law of thermodynamics, to which no violation has hitherto been observed, and the least action principle are valid. The latter, the least action principle, too is a general dynamical law as much as the former, in the sense that from it, such basic mechanical laws as the Euler-Lagrange equations, Newton's second law, the Schr\"odinger equation, and the de Broglie relations (see e.g. \cite{Brizard}), can each be derived.

By its having the dimensions "joule $\times$ second", its universal constancy as mathematically inferred in the foregoing, and its basic relationships with particle dynamical variables as given e.g. by  (\ref{eq-Schra}),  and (\ref{eq-ajmu-a0-2p})--(\ref{eq-bnuxpp}), $\hp_0$ is therefore identifiable with the Planck constant $h$,  $\hp_0=h$.  Accordingly (\ref{eq-Schra}) is identifiable  with the Schr\"odinger equation,  (\ref{eq-ajmu-a0-2p}a)  with 
the Heisenberg uncertainty relations 
$$\displaylines{\refstepcounter{equation} \label{eq-Heis1}
\hfill
\D \P_{\j 1} \D x_{\j 1}=\ldots=\D \P_{\j\n} \D x_{\j\n}
=\D  \Engvar_{\j 1} \D t_{\j 1}=\ldots=
\D  \Engvar_{\j\n} \D t_{\j\n}=h,
\hfill (\ref{eq-Heis1})
}$$
 (\ref{eq-An}) with e.g.  the de Broglie relations $\P_{\j n}\Lam_{\j n} =n h$, $\Engvar_{\j n}\Taum_{\jn}= n h$ (with  $\Ia_{j 0}=0$, $\D X_{\j n}= \Lam_{\j n} $, $\D T_{\j n}=\Taum_{\j n} $, and $n$ effectively continuous) of a free particle,  
and 
(\ref{eq-bnuxpp}) with the usual equation for the smallest  volume $h^{DN}$ accessible to a statistical-mechanical microstate in  phase space.

The author expresses thanks to  Professor Chairman H T Elze
for inviting the author to contribute a talk at the International workshop DICE, Castiglioncello (Tuscany), Italy, Sept, 2012, %Sept 17-21
where the author has enjoyed interesting conversations with several participants,
 to Professor D Schuch for useful discussion  (on the author's another work on emission of radiation quantum) at the workshop, to P-I Johansson for private financial support of the author's research, and to Professors B Johansson,  I Lindgren and others for  giving moral support to the author's research. 
%The author devotes this paper to mother and father.

\end{document}